\documentclass[journal]{IEEEtran}

\usepackage{textcomp}
\usepackage{pgfplots}
\usepackage{amsmath}
\usepackage{multicol}
\usepackage{graphicx}
\graphicspath{{Figures/}}
\usepackage{epstopdf}
\usepackage{acronym}
\usepackage{amsmath}
\usepackage{amsfonts}
\usepackage{color, colortbl}
\usepackage{subfigure}
\usepackage[english]{babel}
\usepackage{blindtext}
\usepackage{algorithm} 
\usepackage{algorithmic} 
\usepackage{multirow}
\usepackage{color}
\usepackage{subfigure}
\usepackage{balance}
\usepackage{caption}

\begin{document}

\newacro{3GPP}{third generation partnership project}
\newacro{4G}{4{th} generation}
\newacro{5G}{5{th} generation}

\newacro{Adam}{adaptive moment estimation}
\newacro{ADC}{analogue-to-digital converter}
\newacro{AED}{accumulated euclidean distance}
\newacro{AGC}{automatic gain control}
\newacro{AI}{artificial intelligence}
\newacro{AMB}{adaptive multi-band}
\newacro{AMB-SEFDM}{adaptive multi-band SEFDM}
\newacro{AN}{artificial noise}
\newacro{ANN}{artificial neural network}
\newacro{AoA}{angle of arrival}
\newacro{ASE}{amplified spontaneous emission}
\newacro{ASIC}{application specific integrated circuit}
\newacro{AWG}{arbitrary waveform generator}
\newacro{AWGN}{additive white Gaussian noise}
\newacro{A/D}{analog-to-digital}

\newacro{B2B}{back-to-back}
\newacro{BCF}{bandwidth compression factor}
\newacro{BCJR}{Bahl-Cocke-Jelinek-Raviv}
\newacro{BDM}{bit division multiplexing}
\newacro{BED}{block efficient detector}
\newacro{BER}{bit error rate}
\newacro{BFW}{beyond-Fourier waveform}
\newacro{BFT}{beyond-Fourier transform}
\newacro{Block-SEFDM}{block-spectrally efficient frequency division multiplexing}
\newacro{BLER}{block error rate}
\newacro{BPSK}{binary phase shift keying}
\newacro{BS}{base station}
\newacro{BSS}{best solution selector}
\newacro{BT}{British Telecom}
\newacro{BU}{butterfly unit}

\newacro{CapEx}{capital expenditure}
\newacro{CA}{carrier aggregation}
\newacro{CBS}{central base station}
\newacro{CC}{component carriers}
\newacro{CCDF}{complementary cumulative distribution function}
\newacro{CCE}{control channel element}
\newacro{CCs}{component carriers}
\newacro{CD}{chromatic dispersion}
\newacro{CDF}{cumulative distribution function}
\newacro{CDI}{channel distortion information}
\newacro{CDMA}{code division multiple access}
\newacro{CDMA2000}{code division multiple access 2000}
\newacro{CI}{constructive interference}
\newacro{CIR}{carrier-to-interference power ratio}
\newacro{CMOS}{complementary metal-oxide-semiconductor}
\newacro{CNN}{convolutional neural network}
\newacro{CoMP}{coordinated multiple point}
\newacro{CO-SEFDM}{coherent optical-SEFDM}
\newacro{CP}{cyclic prefix}
\newacro{CPE}{common phase error}
\newacro{CRVD}{conventional real valued decomposition}
\newacro{CR}{cognitive radio}
\newacro{CRC}{cyclic redundancy check}
\newacro{CS}{central station}
\newacro{CSI}{channel state information}
\newacro{CSIT}{channel state information at the transmitter}
\newacro{CSPR}{carrier to signal power ratio}
\newacro{CW}{continuous-wave}
\newacro{CWT}{continuous wavelet transform}
\newacro{C-RAN}{cloud-radio access networks}

\newacro{DAC}{digital-to-analogue converter}
\newacro{DBP}{digital backward propagation}
\newacro{DC}{direct current}
\newacro{DCGAN}{deep convolutional generative adversarial network}
\newacro{DCI}{downlink control information}
\newacro{DCT}{discrete cosine transform}
\newacro{DDC}{digital down-conversion}
\newacro{DDO-OFDM}{directed detection optical-OFDM}
\newacro{DDO-OFDM}{direct detection optical-OFDM}
\newacro{DDO-SEFDM}{directed detection optical-SEFDM}
\newacro{DFB}{distributed feedback}
\newacro{DFDMA}{distributed FDMA}
\newacro{DFT}{discrete Fourier transform}
\newacro{DFrFT}{discrete fractional Fourier transform}
\newacro{DL}{deep learning}
\newacro{DMA}{direct memory access}
\newacro{DMRS}{demodulation reference signal}
\newacro{DoF}{degree of freedom}
\newacro{DOFDM}{dense orthogonal frequency division multiplexing}
\newacro{DP}{dual polarization}
\newacro{DPC}{dirty paper coding}
\newacro{DSB}{double sideband}
\newacro{DSL}{digital subscriber line}
\newacro{DSP}{digital signal processors}
\newacro{DSSS}{direct sequence spread spectrum}
\newacro{DT}{decision tree}
\newacro{DVB}{digital video broadcast}
\newacro{DWDM}{dense wavelength division multiplexing}
\newacro{DWT}{discrete wavelet transform}
\newacro{D/A}{digital-to-analog}

\newacro{ECC}{error correcting codes}
\newacro{ECL}{external-cavity laser}
\newacro{ECOC}{error-correcting output codes}
\newacro{EDFA}{erbium doped fiber amplifier}
\newacro{EE}{energy efficiency}
\newacro{eMBB}{enhanced mobile broadband}
\newacro{eNB-IoT}{enhanced NB-IoT}
\newacro{EPA}{extended pedestrian A}
\newacro{EVM}{error vector magnitude}

\newacro{Fast-OFDM}{fast-orthogonal frequency division multiplexing}
\newacro{FBMC}{filter bank multicarrier }
\newacro{FCE}{full channel estimation}
\newacro{FD}{fixed detector}
\newacro{FDD}{frequency division duplexing}
\newacro{FDM}{frequency division multiplexing}
\newacro{FDMA}{frequency division multiple access}
\newacro{FE}{full expansion}
\newacro{FEC}{forward error correction}
\newacro{FEXT}{far-end crosstalk}
\newacro{FF}{flip-flop}
\newacro{FFT}{fast Fourier transform}
\newacro{FFTW}{Fastest Fourier Transform in the West}
\newacro{FHSS}{frequency-hopping spread spectrum}
\newacro{FIFO}{first in first out}
\newacro{FMCW}{frequency-modulated continuous wave}
\newacro{F-OFDM}{filtered-orthogonal frequency division multiplexing}
\newacro{FPGA}{field programmable gate array}
\newacro{FrFT}{fractional Fourier transform}
\newacro{FSD}{fixed sphere decoding}
\newacro{FSD-MNSF}{FSD-modified-non-sort-free}
\newacro{FSD-NSF}{FSD-non-sort-free}
\newacro{FSD-SF}{FSD-sort-free}
\newacro{FSK}{frequency shift keying}
\newacro{FTN}{faster than Nyquist}
\newacro{FTTB}{fiber to the building}
\newacro{FTTC}{fiber to the cabinet}
\newacro{FTTdp}{fiber to the distribution point}
\newacro{FTTH}{fiber to the home}

\newacro{GAN}{generative adversarial network}
\newacro{GB}{guard band}
\newacro{GFDM}{generalized frequency division multiplexing}
\newacro{GMSK}{Gaussian minimum-shift keying }
\newacro{GNN}{graph neural networks}
\newacro{GPU}{graphics processing unit}
\newacro{GSM}{global system for mobile communication}
\newacro{GUI}{graphical user interface}

\newacro{HARQ}{hybrid automatic repeat request}
\newacro{HC-MCM}{high compaction multi-carrier communication}
\newacro{HPA}{high power amplifier}

\newacro{IC}{integrated circuit}
\newacro{ICI}{inter carrier interference}
\newacro{ICT}{Information and communications technology}
\newacro{ID}{iterative detection}
\newacro{IDCT}{inverse discrete cosine transform}
\newacro{IDFT}{inverse discrete Fourier transform}
\newacro{IDFrFT}{inverse discrete fractional Fourier transform}
\newacro{ID-FSD}{iterative detection-FSD}
\newacro{ID-SD}{ID-sphere decoding}
\newacro{IF}{intermediate frequency}
\newacro{IFFT}{inverse fast Fourier transform}
\newacro{IFrFT}{inverse fractional Fourier transform}
\newacro{IIoT}{industrial Internet of things}
\newacro{IM}{index modulation}
\newacro{IMD}{intermodulation distortion}
\newacro{INOFS}{inverse non-orthogonal frequency shaping}
\newacro{INOFST}{INOFS transform}
\newacro{IoT}{Internet of things}
\newacro{IOTA}{isotropic orthogonal transform algorithm}
\newacro{IP}{intellectual property}
\newacro{IR}{infrared}
\newacro{ISAC}{integrated sensing and communication}
\newacro{ISAR}{inverse synthetic aperture radar}
\newacro{ISC}{interference self cancellation}
\newacro{ISI}{inter symbol interference}
\newacro{ISM}{industrial, scientific and medical}
\newacro{ISTA}{iterative shrinkage and thresholding}
\newacro{ITU}{international telecommunication union}
\newacro{IUI}{inter user interference}
\newacro{IWAI}{integrated waveform and intelligence}

\newacro{KNN}{k-nearest neighbours}

\newacro{LDPC}{low density parity check}
\newacro{LFDMA}{localized FDMA}
\newacro{LLR}{log-likelihood ratio}
\newacro{LNA}{low noise amplifier}
\newacro{LO}{local oscillator}
\newacro{LOS}{line-of-sight}
\newacro{LPWAN}{low power wide area network}
\newacro{LS}{least square}
\newacro{LSTM}{long short-term memory}
\newacro{LTE}{long term evolution}
\newacro{LTE-Advanced}{long term evolution-advanced}
\newacro{LUT}{look-up table}

\newacro{MA}{multiple access}
\newacro{MAC}{media access control}
\newacro{MAMB}{mixed adaptive multi-band}
\newacro{MAMB-SEFDM}{mixed adaptive multi-band SEFDM}
\newacro{MASK}{m-ary amplitude shift keying}
\newacro{MB}{multi-band}
\newacro{MB-SEFDM}{multi-band SEFDM}
\newacro{MCM}{multi-carrier modulation}
\newacro{MC-CDMA}{multi-carrier code division multiple access}
\newacro{MCS}{modulation and coding scheme}
\newacro{MF}{matched filter}
\newacro{MIMO}{multiple input multiple output}
\newacro{ML}{maximum likelihood}
\newacro{MLSD}{maximum likelihood sequence detection}
\newacro{MMF}{multi-mode fiber}
\newacro{MMSE}{minimum mean squared error}
\newacro{mMTC}{massive machine-type communication}
\newacro{MNSF}{modified-non-sort-free}
\newacro{MOFDM}{masked-OFDM}
\newacro{MRVD}{modified real valued decomposition}
\newacro{MS}{mobile station}
\newacro{MSE}{mean squared error}
\newacro{MTC}{machine-type communication}
\newacro{MUI}{multi-user interference}
\newacro{MUSA}{multi-user shared access}
\newacro{MU-MIMO}{multi-user multiple-input multiple-output}
\newacro{MZM}{Mach-Zehnder modulator}
\newacro{M2M}{machine to machine}

\newacro{NB-IoT}{narrowband IoT}
\newacro{NB}{naive Bayesian}
\newacro{NDFF}{National Dark Fiber Facility}
\newacro{NEXT}{near-end crosstalk}
\newacro{NFV}{network function virtualization}
\newacro{NG-IoT}{next generation IoT}
\newacro{NLOS}{non-line-of-sight}
\newacro{NLSE}{nonlinear Schrödinger equation}
\newacro{NN}{neural network}
\newacro{NOFDM}{non-orthogonal frequency division multiplexing}
\newacro{NOMA}{non-orthogonal multiple access}
\newacro{NoFDMA}{non-orthogonal frequency division multiple access}
\newacro{NOFS}{non-orthogonal frequency spacing}
\newacro{NOFST}{NOFS transform}
\newacro{NP}{non-polynomial}
\newacro{NR}{new radio}
\newacro{NSF}{non-sort-free}
\newacro{NWDM}{Nyquist wavelength division multiplexing }
\newacro{Nyquist-SEFDM}{Nyquist-spectrally efficient frequency division multiplexing}

\newacro{OBM-OFDM}{orthogonal band multiplexed OFDM}
\newacro{ODDM}{orthogonal delay-Doppler division multiplexing}
\newacro{OF}{optical filter}
\newacro{OFDM}{orthogonal frequency division multiplexing}
\newacro{OFDMA}{orthogonal frequency division multiple access}
\newacro{OMA}{orthogonal multiple access}
\newacro{OpEx}{operating expenditure}
\newacro{OPM}{optical performance monitoring}
\newacro{OQAM}{offset-QAM}
\newacro{OSI}{open systems interconnection}
\newacro{OSNR}{optical signal-to-noise ratio}
\newacro{OSSB}{optical single sideband}
\newacro{OTA}{over-the-air}
\newacro{OTFS}{orthogonal time frequency space}
\newacro{Ov-FDM}{Overlapped FDM}
\newacro{O-SEFDM}{optical-spectrally efficient frequency division multiplexing}
\newacro{O-FOFDM}{optical-fast orthogonal frequency division multiplexing}
\newacro{O-OFDM}{optical-orthogonal frequency division multiplexing}
\newacro{O-CDMA}{optical-code division multiple access}

\newacro{PA}{power amplifier}
\newacro{PAPR}{peak-to-average power ratio}
\newacro{PCA}{principal component analysis}
\newacro{PCE}{partial channel estimation}
\newacro{PD}{photodiode}
\newacro{PDCCH}{physical downlink control channel}
\newacro{PDF}{probability density function}
\newacro{PDP}{power delay profile}
\newacro{PDMA}{polarisation division multiple access}
\newacro{PDM-OFDM}{polarization-division multiplexing-OFDM}
\newacro{PDM-SEFDM}{polarization-division multiplexing-SEFDM}
\newacro{PDSCH}{physical downlink shared channel}
\newacro{PE}{processing element}
\newacro{PED}{partial Euclidean distance}
\newacro{PHY}{physical-layer}
\newacro{PLA}{physical layer authentication}
\newacro{PLS}{physical layer security}
\newacro{PMD}{polarization mode dispersion}
\newacro{PON}{passive optical network}
\newacro{PPM}{parts per million}
\newacro{PRB}{physical resource block}
\newacro{PSD}{power spectral density}
\newacro{PSK}{pre-shared key}
\newacro{PSNR}{peak signal-to-noise ratio}
\newacro{PSS}{primary synchronization signal}
\newacro{PU}{primary user}
\newacro{PXI}{PCI extensions for instrumentation}
\newacro{P/S}{parallel-to-serial}

\newacro{QAM}{quadrature amplitude modulation}
\newacro{QKD}{quantum key distribution}
\newacro{QoS}{quality of service}
\newacro{QPSK}{quadrature phase-shift keying}
\newacro{QRNG}{quantum random number generation}

\newacro{RAUs}{remote antenna units}
\newacro{RAT}{radio access technology}
\newacro{RBF}{radial basis function}
\newacro{RBW}{resolution bandwidth}
\newacro{ReLU}{rectified linear units}
\newacro{RF}{radio frequency}
\newacro{RMS}{root mean square}
\newacro{RMSE}{root mean square error}
\newacro{RMSProp}{root mean square propagation}
\newacro{RNTI}{radio network temporary identifier}
\newacro{RoF}{radio-over-fiber}
\newacro{ROM}{read only memory}
\newacro{RRC}{root raised cosine}
\newacro{RC}{raised cosine}
\newacro{RSC}{recursive systematic convolutional}
\newacro{RSSI}{received signal strength indicator}
\newacro{RTL}{register transfer level}
\newacro{RVD}{real valued decomposition}

\newacro{SB-SEFDM}{single-band SEFDM}
\newacro{ScIR}{sub-carrier to interference ratio}
\newacro{SCMA}{sparse code multiple access}
\newacro{SC-NOFS}{single-carrier non-orthogonal frequency shaping}
\newacro{SC-OFDM}{single-carrier orthogonal frequency division multiplexing}
\newacro{SC-FDMA}{single-carrier frequency division multiple access}
\newacro{SC-SEFDMA}{single-carrier spectrally efficient frequency division multiple access}
\newacro{SD}{sphere decoding}
\newacro{SDM}{space division multiplexing}
\newacro{SDMA}{space division multiple access}
\newacro{SDN}{software-defined network}
\newacro{SDP}{semidefinite programming}
\newacro{SDR}{software-defined radio}
\newacro{SE}{spectral efficiency}
\newacro{SEFDM}{spectrally efficient frequency division multiplexing}
\newacro{SEFDMA}{spectrally efficient frequency division multiple access} 
\newacro{SF}{sort-free}
\newacro{SFCW}{stepped-frequency continuous wave}
\newacro{SGD}{stochastic gradient descent}
\newacro{SGDM}{stochastic gradient descent with momentum}
\newacro{SIC}{successive interference cancellation}
\newacro{SiGe}{silicon-germanium}
\newacro{SINR}{signal-to-interference-plus-noise ratio}
\newacro{SIR}{signal-to-interference ratio}
\newacro{SISO}{single-input single-output}
\newacro{SLM}{spatial light modulator}
\newacro{SMF}{single mode fiber}
\newacro{SNR}{signal-to-noise ratio}
\newacro{SP}{shortest-path}
\newacro{SPSC}{symbol per signal class}
\newacro{SPM}{self-phase modulation}
\newacro{SRS}{sounding reference signal}
\newacro{SSB}{single-sideband}
\newacro{SSBI}{signal-signal beat interference}
\newacro{SSFM}{split-step Fourier method}
\newacro{SSMF}{standard single mode fiber}
\newacro{STBC}{space time block coding}
\newacro{STFT}{short time Fourier transform}
\newacro{STC}{space time coding}
\newacro{STO}{symbol timing offset}
\newacro{SU}{secondary user}
\newacro{SVD}{singular value decomposition}
\newacro{SVM}{support vector machine}
\newacro{SVR}{singular value reconstruction}
\newacro{S/P}{serial-to-parallel}

\newacro{TDD}{time division duplexing}
\newacro{TDMA}{time division multiple access }
\newacro{TDM}{time division multiplexing}
\newacro{TFP}{time frequency packing}
\newacro{THP}{Tomlinson-Harashima precoding}
\newacro{TOFDM}{truncated OFDM}
\newacro{TSPSC}{training symbols per signal class}
\newacro{TSVD}{truncated singular value decomposition}
\newacro{TSVD-FSD}{truncated singular value decomposition-fixed sphere decoding}
\newacro{TTI}{transmission time interval}

\newacro{UAV}{unmanned aerial vehicle}
\newacro{UCR}{user compression ratio}
\newacro{UE}{user equipment}
\newacro{UFMC}{universal-filtered multi-carrier}
\newacro{ULA}{uniform linear array}
\newacro{UMTS}{universal mobile telecommunications service}
\newacro{URLLC}{ultra-reliable low-latency communication}
\newacro{USRP}{universal software radio peripheral}
\newacro{UWB}{ultra-wideband}

\newacro{VDSL}{very-high-bit-rate digital subscriber line}
\newacro{VDSL2}{very-high-bit-rate digital subscriber line 2}
\newacro{VHDL}{very high speed integrated circuit hardware description language}
\newacro{VLC}{visible light communication}
\newacro{VLSI}{very large scale integration}
\newacro{VOA}{variable optical attenuator}
\newacro{VP}{vector perturbation}
\newacro{VSSB-OFDM}{virtual single-sideband OFDM}
\newacro{V2V}{vehicle-to-vehicle}

\newacro{WAN}{wide area network}
\newacro{WCDMA}{wideband code division multiple access}
\newacro{WDM}{wavelength division multiplexing}
\newacro{WDP}{waveform-defined privacy}
\newacro{WDS}{waveform-defined security}
\newacro{WiFi}{wireless fidelity}
\newacro{WiGig}{Wireless Gigabit Alliance}
\newacro{WiMAX}{Worldwide interoperability for Microwave Access}
\newacro{WLAN}{wireless local area network}
\newacro{WSS}{wavelength selective switch}

\newacro{XPM}{cross-phase modulation}

\newacro{ZF}{zero forcing}
\newacro{ZP}{zero padding}


\title{Rethinking Next-Generation Signal Waveform: Integration of Orthogonality and Non-Orthogonality}

\author{{Tongyang Xu,~\IEEEmembership{Senior Member,~IEEE}, Shuangyang Li,~\IEEEmembership{Member,~IEEE}, Zhongxiang Wei,~\IEEEmembership{Member,~IEEE},\\ Gan Zheng,~\IEEEmembership{Fellow,~IEEE}, Izzat Darwazeh,~\IEEEmembership{Senior Member,~IEEE}}

\thanks{
This work was supported in part by UK Engineering and Physical Sciences Research Council (EPSRC) IWAI under Grant EP/Y000315/2; in part by EPSRC TRACCS under Grant EP/W026252/1; in part by the UK EPSRC under Grant EP/X04047X/2 and Grant EP/Y037243/1; in part by the Fundamental Research Funds for the Central Universities under Grant 22120230311; in part by Shanghai Municipal Science and Technology Program Projects under Grant 25ZR1401361; in part by the National Key Laboratory of Avionics Integration and Aviation System-of-Systems Synthesis under grant 2025AIASS0701; and in part by the European Research Council (ERC) under the ERC Starting Grant 101220383 (FUNDOCS).

T. Xu and I. Darwazeh are with the Department of Electronic and Electrical Engineering, University College London, WC1E 7JE, U.K. (e-mail: tongyang.xu.11@ucl.ac.uk, i.darwazeh@ucl.ac.uk).

S. Li is with the Department of Electrical Engineering and Computer Science, Technical University of Berlin, Berlin 10587, Germany. (e-mail: shuangyang.li@tuberlin.de).

Z. Wei is with the School of Electronics and Information Engineering, Tongji University, Shanghai 200070, China. (e-mail: z\_wei@tongji.edu.cn).

G. Zheng is with the School of Engineering, The University of Warwick, CV4 7AL, Coventry, U.K. (e-mail: gan.zheng@warwick.ac.uk).
}
}

\maketitle

\begin{abstract}

As 6G communications advance, the demand for new services and capabilities, as defined by the international telecommunication union (ITU), is increasing. A crucial aspect of 6G advancement lies in the development of signal waveforms that can meet these demands while maintaining compatibility with existing standards. This paper explores sustainable physical layer waveform options, focusing on a balanced approach that integrates non-orthogonality with orthogonality to achieve both backward compatibility and forward innovation. Specifically, we investigate two key signal formats: single-carrier orthogonal frequency division multiplexing (SC-OFDM) (1D,2D) and single-carrier non-orthogonal frequency shaping (SC-NOFS)(1D,2D). Both can use 1D frequency and 2D time-frequency precoding, offering enhanced frequency and time diversity, simplified processing, and resilience to delay-Doppler effects. SC-NOFS(2D) further introduces advantages such as improved spectral efficiency and reduced latency, making it a strong candidate for future 6G applications. The comparative analysis highlights that SC-NOFS(2D) provides a broader range of capabilities,  particularly those requiring high data rate, high mobility, low-latency communication, sustainability, and interoperability, positioning it as a versatile solution for next-generation 6G communication.

\end{abstract}


\begin{figure*}[t]
\begin{center}
\includegraphics[scale=0.435]{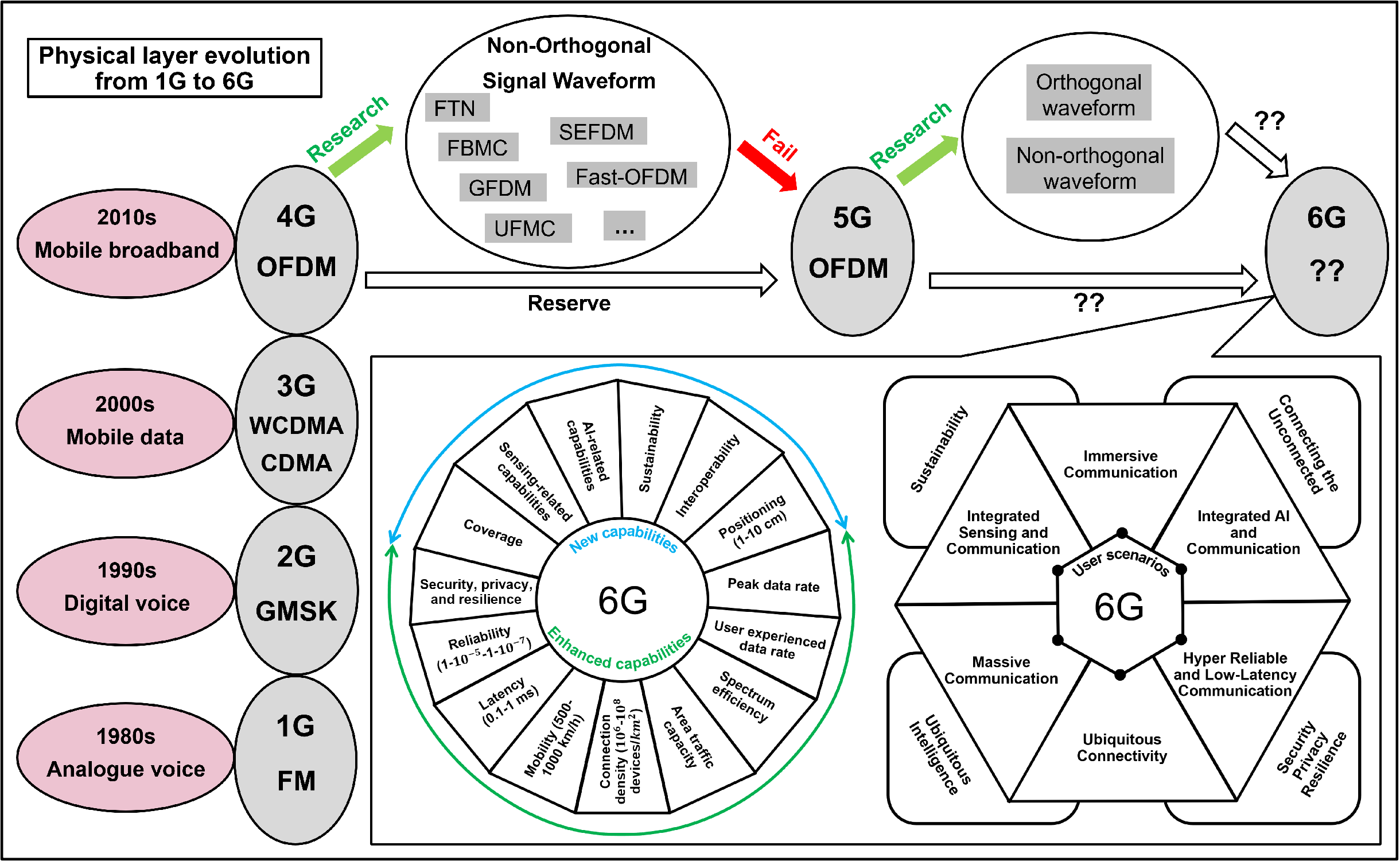}
\end{center}
\caption{Physical layer waveform innovations from 1G to future 6G communications.}
\label{Fig:SC_NOFS_1G_6G_development}
\end{figure*}

\section{Introduction}

The development of communication techniques is ongoing with innovative waveform advancement in each generation as illustrated {in Fig. \ref{Fig:SC_NOFS_1G_6G_development}. The first generation (1G), appeared in 1980s, relied on analog modulation techniques, frequency modulation (FM), for voice communications. However, due to the low-quality of voice and no support for data transfer, 1G was replaced by the second generation (2G) in 1990s when digital communications started to dominate. The representative in 2G is \ac{GSM} where digital modulation techniques, \ac{GMSK}, is developed. With the digital technology, voice quality is greatly improved. In addition, new services such as data transfer and text messaging were introduced. However, data rate was still very low and data roaming was not supported. In early 2000s, the third generation (3G) communications, using \ac{UMTS} and \ac{CDMA2000}, introduced higher data rates, mobile internet access, video calling, and multimedia services. These systems rely on \ac{WCDMA} and CDMA techniques. To further improve data rates and offer more robust services, in 2010s, the fourth generation (4G) communication, \ac{LTE}, is introduced with a significant technique breakthrough in utilizing \ac{OFDM} physical layer waveforms. OFDM represents a transformative leap and brings new features to communications over other conventional waveform solutions. Firstly, it is robust to multipath frequency selective channels; then it has higher spectral efficiency; moreover, its signal processing is relatively simple. With a multi-carrier signal format, the 4G system can provide significantly faster data rates, low-latency communication, enhanced multimedia services, and better smartphone user experience. Due to the huge success in 4G and particularly the \ac{OFDM} waveform, many research groups focused on new signal waveform design in early 2010s, for the purpose of showing impact in the fifth generation (5G).}

As illustrated in Fig. \ref{Fig:SC_NOFS_1G_6G_development} that many waveform candidates were explored for 5G such as \ac{FTN} \cite{Anderson2013}, \ac{FBMC} \cite{FBMC2011_magazine}, \ac{GFDM} \cite{GFDM_trans}, \ac{UFMC} \cite{UFMC2013}, \ac{SEFDM} \cite{TongyangTVT2017}, \ac{Fast-OFDM} \cite{Tongyang_JIOT_2018_double_device}. All of these candidates are grouped in non-orthogonal waveforms with advantages such as faster data rate, reduced out-of-band power leakage, and reduced spectral bandwidth utilization. However, even with the significant benefits obtained from the above non-orthogonal signal waveforms, none of them is really considered for 5G \cite{Erik_book_5G}. One critical reason is that those non-orthogonal signal waveforms obtain enhanced spectral efficiency but at the cost of increased computational complexity. The extra power consumption caused by the increased computational complexity compromizes the increased spectral efficiency achieved by non-orthogonal waveforms. Moreover, operators have invested hugely on telecommunication infrastructure. The change of physical layer waveform requires them to replace old hardware infrastructure, which is economically inefficient. Due to these reasons, 5G still reused OFDM as its physical layer signal waveform with more flexible configurations to support different services such as \ac{eMBB}, \ac{URLLC}, and \ac{mMTC}.

For future 6G, more services are expected by operators \cite{6G_operator_CM_2024} and more capabilities are coming out as defined by \ac{ITU} \cite{6G_IMT2030}. The significance of signal waveform for future 6G communications is apparent and the way to move forward signal waveforms to 6G is still under research in terms of orthogonal and non-orthogonal waveforms. In envisioning the signal waveform for 6G, our perspective emphasizes a design that not only introduces novel capabilities but also maintains compatibility with existing standards. The aim is to encourage innovation and advancement while ensuring a seamless integration with the established frameworks of previous generations. This vision states the importance of evolutionary continuity, allowing for the coexistence of emerging technologies with the infrastructure of preceding wireless communication standards. To show the integration of innovation and interoperability, this work will introduce potential waveform candidates, including \ac{OFDM}, \ac{NOFS} \cite{Tongyang_Nature_2023}, \ac{SC-OFDM}(1D), \ac{SC-NOFS}(1D) \cite{Tongyang_JIOT_2024_SC_OFDM}, \ac{SC-OFDM}(2D), and \ac{SC-NOFS}(2D). Among these, SC-NOFS(2D) stands out as a particularly promising candidate for 6G, offering significant advantages in terms of spectral efficiency, robustness to delay-Doppler, and future scalability while maintaining compatibility with existing OFDM-based systems.

\begin{figure*}[t!]
\begin{center}
\includegraphics[scale=0.68]{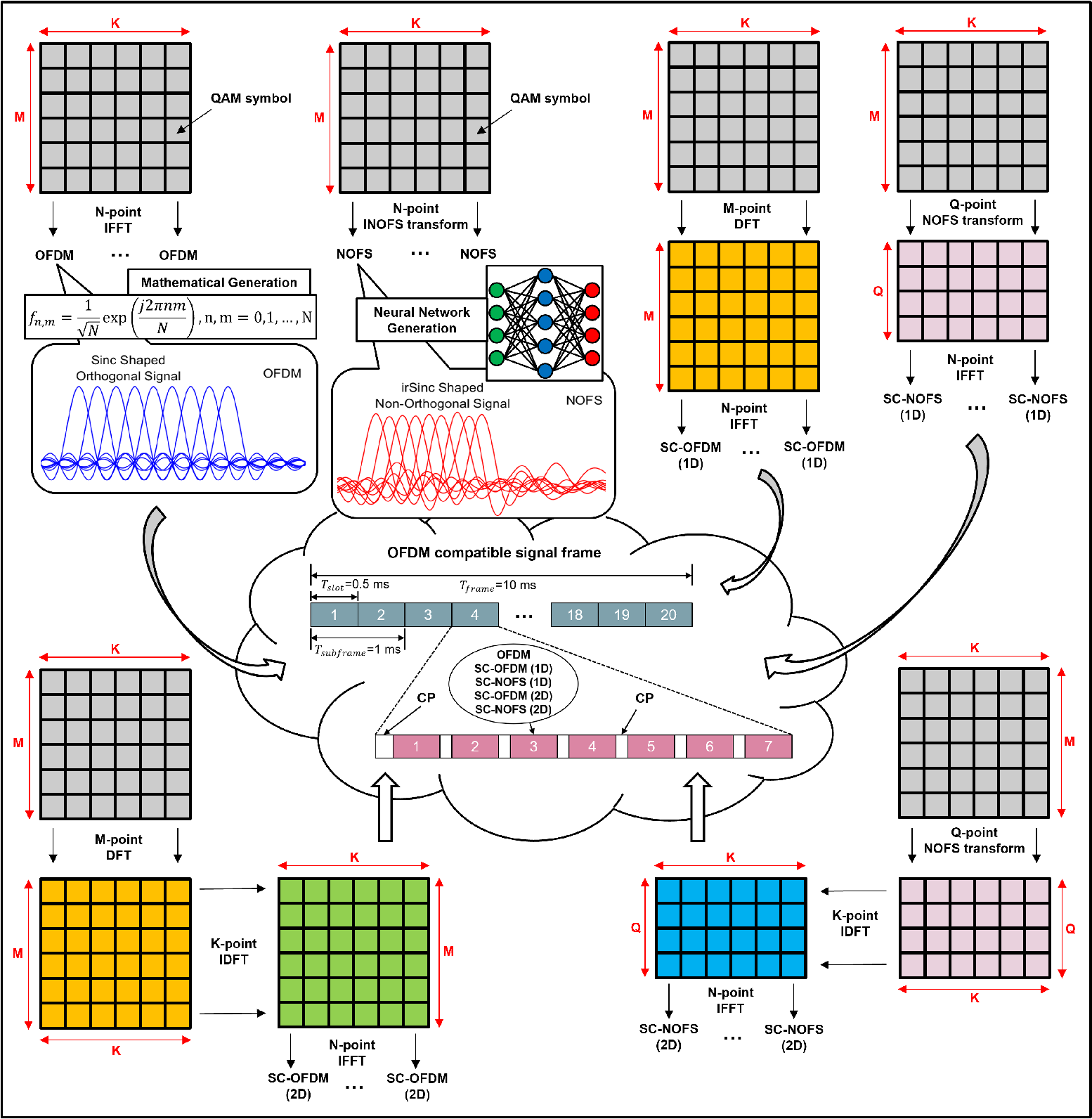}
\end{center}
\caption{Principle of orthogonal waveforms (OFDM, SC-OFDM(1D), SC-OFDM(2D)) and non-orthogonal waveforms (NOFS, SC-NOFS(1D), SC-NOFS(2D)). {Orthogonal OFDM-based waveforms are generated using deterministic mathematical models, whereas non-orthogonal NOFS-based waveforms are generated via neural networks, with each generation coefficient optimized through machine learning.} }
\label{Fig:SC_NOFS_comparison_other_signals}
\end{figure*}

\section{Motivation}
\subsection{Problem Statement}

In 4G and 5G communication systems, OFDM and SC-OFDM are the standardized physical layer waveforms. One crucial consideration for such waveforms is their low complexity where both OFDM and SC-OFDM are generated via \ac{IFFT}. Another important concern is their straightforward channel estimation and equalization. OFDM is an orthogonal signal waveform where its channel estimation and equalization can be realized via the frequency-domain one-tap processing. In addition, its multiple sub-carrier signal structure provides robustness to multipath frequency selective channel effects. With the single carrier capability, SC-OFDM offers additional low \ac{PAPR} benefits. Moreover, the signal structure of OFDM can be easily integrated in \ac{MIMO} systems.

Advanced signal waveforms were proposed for 5G when OFDM had not been confirmed to be the physical layer waveform standard. All the signal waveforms bring new features over OFDM such as higher data rate, reduced spectral utilization, and low out-of-band power leakage. However, their signal structures are quite different compared with OFDM, which is indeed one critical issue because operators have invested a lot in 4G where OFDM has been integrated in their infrastructure. Therefore, maintaining OFDM waveform while introducing additional functions would be a desired option. Moreover, those new signal waveforms are all non-orthogonally structured, which breaks the orthogonality and complicates channel estimation and equalization. Traditional channel estimation algorithms rely on one-tap signal processing while with the non-orthogonality channel estimation has to be managed in time-domain or frequency-domain with \ac{ICI}. It is noted that time-domain channel signal processing involves a large matrix inverse and when the matrix is not well structured, the inverse will be inaccurate. Therefore, the question is whether it is worth using non-orthogonal signals with higher spectral efficiency but at the cost of complex signal processing or even inaccurate processing.

\subsection{Sustainability and Interoperability for 6G} \label{subsec:Sustainability_and_Interoperability}

\ac{ITU} \cite{6G_IMT2030} has defined 6G user scenarios, new capabilities, and enhanced capabilities in Fig. \ref{Fig:SC_NOFS_1G_6G_development}. {It is apparent that the critical factor, both mentioned in application scenarios and new capabilities, is sustainability \cite{GreenCom_TGCN_2022}. Sustainability refers to the ability of a waveform to maintain compatibility with existing communication infrastructure, especially the reuse of over-the-air interfaces and baseband processing hardware, and therefore this reduces the costs of adopting new waveforms. It is anticipated that future 6G communications would make use of available core hardware infrastructure and only update low-cost hardware components such as antennas and analogue components that can support high frequency range. In addition, interoperability across different generations is another new capability introduced in 6G, emphasizing the importance of compatibility between techniques. Both the sustainability and interoperability are crucial factors that greatly affect the 6G investment by operators \cite{6G_operator_CM_2024,6G_green_CM_2023}. Operators have invested a lot to legacy communication systems and they need time to pay back. Therefore, significant changes of existing hardware due to new technology development would be unacceptable.}

Sustainability and interoperability in 6G are important, but new services and new capabilities are still needed to meet future demands and should be integrated in 6G. The current 5G standard utilizes the \ac{OFDM} waveform, employing orthogonally packed sub-carriers that facilitate simple signal generation, channel estimation, channel equalization, and signal detection. To enhance the physical layer signal waveform performance and bring new capabilities and user application scenarios, non-orthogonality emerges as a key innovation. Non-orthogonal waveforms can significantly enhance spectral efficiency, unlocking new possibilities for future communication systems. However, these waveforms also introduce challenges, including increased complexity in signal generation, channel estimation, channel equalization, and signal detection. Furthermore, non-orthogonality complicates signal integration in \ac{MIMO} systems due to inherent signal interference.

Therefore, maintaining signal orthogonality over the air is crucial for achieving low-complexity signal processing, thereby supporting the sustainability objective. Here, we consider waveform candidates that are compatible with the 5G OFDM standard, excluding alternatives that require modifications to the OFDM over-the-air interface, even if they promise advanced capabilities. We will discuss some potential signal waveforms and further explain how we can integrate non-orthogonality advantages in orthogonal signals to show the interoperability. This paper explores potential waveform candidates and demonstrates how non-orthogonality can be integrated into orthogonal signals, preserving backward compatibility while achieving forward innovation for 6G.

In this work, we will focus on three key objectives:

\begin{itemize} 

\item{Physical Layer Signal Waveform Backward Compatibility: The over-the-air waveform should be based on OFDM with orthogonally packed sub-carriers. This approach simplifies channel estimation and equalization processes, ensuring seamless integration with existing systems. }  

\item{Signal Frame Structure Backward Compatibility: The signal frame design should align with 5G standards, considering factors such as coherence time, coherence bandwidth, \ac{CP}, and pilot allocation, maintaining consistency with the 5G physical layer frame structure.}

\item{Enhanced Forward Capabilities: Building on the 5G standard compatibility in both the physical layer waveform and signal frame structure, the system should introduce additional benefits, including improved spectral efficiency, higher data rates, or reduced resource utilization.}

\end{itemize}

\section{Physical Layer Waveform Potentials for 6G}\label{sec:potential_6G_waveforms}

\begin{table*}[t!]
\begin{center}
\includegraphics[scale=1.55]{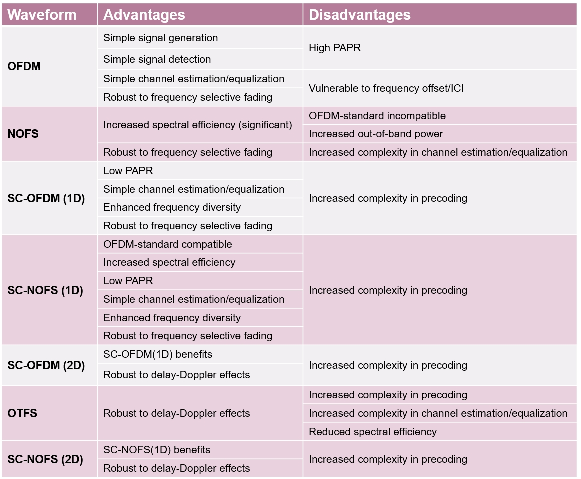}
\end{center}
\caption{{Advantages and disadvantages of orthogonal waveforms (OFDM, SC-OFDM(1D), SC-OFDM(2D), OTFS) and non-orthogonal waveforms (NOFS, SC-NOFS(1D), SC-NOFS(2D)).} }
\label{Fig:SC_NOFS_2D_signal_waveform_adv_disadv}
\end{table*}

This section aims to compare different waveform candidates in terms of signal principle in Fig. \ref{Fig:SC_NOFS_comparison_other_signals}, and advantages and disadvantages in Table \ref{Fig:SC_NOFS_2D_signal_waveform_adv_disadv}.

\subsection{OFDM}

OFDM is the signal waveform standardized in 4G, 5G, and \ac{WLAN} systems. Its signal principle is demonstrated in Fig. \ref{Fig:SC_NOFS_comparison_other_signals}. First, a two-dimensional time-frequency grid $\mathbf{X}\in\mathbb{C}^{M{\times}K}$ is created, where $M$ indicates the number of QAM symbols modulated in one OFDM symbol and $K$ represents the number of OFDM symbols. Each column operates an $N$-point \ac{IFFT} to generate an \ac{OFDM} symbol. The resulting $\mathbf{Y}\in\mathbb{C}^{N{\times}K}$ denotes the OFDM signal matrix consisting of $K$ OFDM symbols, each containing $N$ samples where $M<N$. {The OFDM waveform is generated using a deterministic mathematical model based on \ac{IFFT} coefficients $f_{n,m}$, which enables a Sinc-shaped orthogonal sub-carrier packing scheme illustrated in Fig. \ref{Fig:SC_NOFS_comparison_other_signals}. All OFDM-based waveforms discussed in the following rely on such deterministic models.} The Sinc-shaped orthogonal sub-carrier packing highlights several key advantages of OFDM. First, its spectral efficiency is enhanced by the orthogonal packing of overlapping sub-carriers. Second, its multicarrier structure is resilient to multipath frequency-selective channels, as the channel response on each sub-carrier approaches a constant. Additionally, the insertion of a \ac{CP} at the beginning of each OFDM symbol mitigates \ac{ISI} and simplifies channel estimation and equalization. OFDM signal processing is also simplified by the use of FFT and IFFT operations. Another benefit is its resistance to narrowband interference, as interference typically affects only one sub-carrier, while the remaining sub-carriers continue to transmit data. Moreover, OFDM supports multi-user applications, where each user occupies a different set of sub-carriers, ensuring orthogonal packing without interference. A unique feature of OFDM is its support for adaptive modulation, allowing different modulation schemes and coding rates to be applied to individual sub-carriers based on channel quality. For poor-quality channels, low-order modulation and coding rates can be applied, while high-order modulation and coding rates can be used for better-quality channels.

However, OFDM has several drawbacks. One significant issue is its high \ac{PAPR}, resulting from the summation of multiple sub-carriers, which creates high peak power levels. Additionally, OFDM is highly sensitive to frequency offsets, as maintaining sub-carrier orthogonality is essential. Any disruption in orthogonality can lead to \ac{ICI}, significantly degrading system performance.

\subsection{NOFS}

Recently, a novel non-orthogonal signal format termed \ac{NOFS} \cite{Tongyang_Nature_2023} has been introduced, replacing traditional Sinc-shaped sub-carriers with irregular Sinc (irSinc) sub-carriers, as shown in Fig. \ref{Fig:SC_NOFS_comparison_other_signals}. While the irSinc shaped sub-carriers cause \ac{ICI} in orthogonal OFDM systems, it benefits non-orthogonal signals like NOFS, where sub-carriers are intentionally packed non-orthogonally leading to increased spectral efficiency. The irregular Sinc shapes help mitigate the ICI caused by non-orthogonal packing. Research has shown that NOFS is capable of achieving high spectral efficiency with these irregular Sinc sub-carriers, offering a significant advantage by compressing up to 60\% of spectral resources without sacrificing system performance \cite{Tongyang_Nature_2023}.

Unlike OFDM-based waveforms that rely on deterministic mathematical models, NOFS-based waveforms are generated through data-driven machine learning-optimized neural networks, as shown in Fig. \ref{Fig:SC_NOFS_comparison_other_signals}. The generation process employs a densely connected neural architecture where connection weights determine neuron importance and the network size determines signal generation accuracy. To enable non-orthogonal subcarrier shaping in NOFS, we introduce two key transforms: the $N$-point \ac{NOFST} and its inverse, $N$-point \ac{INOFST}. The \ac{INOFST} replaces conventional fixed IFFT signal generation coefficients with a spectral shaping process that maps input QAM symbols onto a set of irSinc basis functions. These basis functions allow for non-orthogonal packing of subcarriers, enabling spectral compression and increased spectral efficiency. At the receiver, the NOFST is used to reconstruct the transmitted signal while accounting for the controlled self-interference introduced by the non-orthogonal shaping. The key innovation of the NOFS signal, compared to the OFDM signal, lies in its adjustable signal generation parameters. The NOFS-based transforms leverage machine learning to adaptively optimize the signal generation coefficients, effectively mitigating interference from non-orthogonal signals. These signal generation coefficients can be dynamically tuned using predefined or learned transform matrices, as detailed in our previous work \cite{Tongyang_Nature_2023}. This adaptability avoids the fixed IFFT coefficients used in OFDM and explains why the sub-carriers in NOFS, as illustrated in Fig. \ref{Fig:SC_NOFS_comparison_other_signals}, exhibit irSinc-shaped waveforms with distinct patterns.

However, there are several drawbacks with NOFS. The use of irregular Sinc sub-carriers leads to high out-of-band power leakage, as seen in the packing scheme in Fig. \ref{Fig:SC_NOFS_comparison_other_signals}. Another challenge is the complexity of channel estimation, which requires a two-dimensional time-domain estimation method rather than the simpler one-tap frequency-domain approach. Additionally, NOFS is a new waveform design that is not compatible with existing OFDM standards.

\subsection{SC-OFDM(1D)}

SC-OFDM(1D) is a single-carrier variation of OFDM that introduces a precoding stage before the IFFT, often referred to as DFT-spread OFDM because this precoding is performed using a \ac{DFT}, as shown in Fig. \ref{Fig:SC_NOFS_comparison_other_signals}. The two-stage signal generation includes a first-stage $M$-point \ac{DFT} operation for $K$ columns and the second-stage $N$-point \ac{IFFT} operation for each column, where $M<N$. The SC-NOFS(1D) signal is so named because it has one precoding stage and its precoding focuses only on frequency diversity enhancement.

One of the primary advantages of SC-OFDM(1D) is its reduced \ac{PAPR}, achieved through DFT spreading, where symbols are distributed across all sub-carriers. This reduction in PAPR is the main reason why SC-OFDM is utilized in uplink channels for 4G and 5G, while traditional OFDM is used for downlink channels. In addition to reduced PAPR, SC-OFDM also provides enhanced frequency diversity due to the DFT precoding. This frequency diversity allows SC-OFDM to perform better than OFDM in multipath frequency-selective channels. The main drawback of SC-OFDM, however, is the slightly increased signal processing complexity caused by the additional precoding stage before the IFFT.

\subsection{SC-NOFS(1D)}

To address the out-of-band power leakage and signal processing complexity in NOFS signals, an alternative waveform structure based on a single-carrier signal design is proposed, as shown in Fig. \ref{Fig:SC_NOFS_comparison_other_signals}. Similar to SC-OFDM(1D), the two-stage signal generation for SC-NOFS(1D) begins with a precoding stage, where a $Q$-point \ac{NOFST} is applied. The primary function of \ac{NOFST} is to compress the $M$-length QAM vector into a shorter $Q$-length vector, reducing the number of samples passed to the second stage. This compression not only improves spectral efficiency \cite{Tongyang_JIOT_2024_SC_OFDM} over SC-OFDM(1D) but also integrates the non-orthogonal NOFS principle. The shorter sequence generated during this stage is then passed to a second $N$-point IFFT block, producing OFDM-compatible signals, where $Q < M < N$. Despite the signal truncation that introduces interference, the irSinc-shaped transform in \ac{NOFST} ensures perfect interference removal and maintains performance. This improved waveform structure achieves enhanced spectral efficiency while retaining the frequency diversity of conventional SC-OFDM(1D), as the signal spreading in the first stage ensures diversity. Similar to SC-OFDM(1D), SC-NOFS(1D) involves a trade-off with a slight increase in signal processing complexity compared to OFDM due to the use of the NOFS transform. However, it is important to note that the complexity of the NOFS transform in SC-NOFS(1D) is simpler than the DFT used in SC-OFDM(1D), as demonstrated in our previous work \cite{Tongyang_JIOT_2024_SC_OFDM}. {It is important to note that another non-orthogonal technique, known as \ac{NOMA}, is designed for multiple user access, whereas SC-NOFS is a physical-layer waveform design. In fact, SC-NOFS can be used as an underlying waveform within NOMA frameworks, enabling additional gains in spectral efficiency.}

\subsection{SC-OFDM(2D)}

The traditional approach to managing delay-Doppler effects in 5G involves adjusting sub-carrier spacing, data bandwidth, and carrier frequency. This adaptive method is simple, effective, and easily reconfigurable in practical communication systems. Alternatively, new signal waveforms can be designed to improve performance under strong delay-Doppler conditions. One candidate is \ac{OTFS} \cite{OTFS_WCNC2017}, which uses 2D precoding to transform a signal from the delay-Doppler (DD) domain to the time-frequency (TF) domain. However, OTFS has many implementation challenges including signal processing complexity, channel estimation difficulties, and restricted application scenarios. To avoid channel interference, protection gaps are reserved around OTFS pilot symbols, which is a waste of resource, making OTFS challenging to integrate within an OFDM-based signal framework. It is noted that the key objective of this work is to evaluate the performance of various signal waveforms in 3GPP-standard scenarios, specifically those relying on the OFDM waveform format for over-the-air communication. In addition, this work focuses on signal frame backward compatibility where the frame structure has to consider coherence time and coherence bandwidth to align with 5G physical layer frame structure as illustrated in Fig. \ref{Fig:SC_NOFS_comparison_other_signals}. To address multipath fading, each multicarrier symbol must include a \ac{CP} to ensure signal integrity. The delay spread must be contained within the CP duration, and channel estimation must occur within the channel's coherence time. Otherwise, the widely used one-tap frequency-domain channel estimation and equalization techniques will fail.

Due to the distinct requirements for channel estimation, equalization, and pilot allocation in OTFS \cite{OTFS_WCNC2017}, achieving a fair comparison with other standard-compatible waveforms is challenging. Consequently, this work excludes OTFS and instead proposes a similar yet simpler signal format: SC-OFDM(2D). As depicted in Fig. \ref{Fig:SC_NOFS_comparison_other_signals}, SC-OFDM(2D) employs a two-stage precoding process similar to OTFS. Both SC-OFDM(2D) and OTFS extend the conventional SC-OFDM(1D), by introducing an additional IDFT operation to enable two-dimensional signal processing. While OTFS employs unique pilot structures and CP configurations, the proposed SC-OFDM (2D) strictly follows 3GPP specifications with standard-compliant pilot and CP designs. Hence, despite sharing a common foundation in SC-OFDM(1D), SC-OFDM(2D) and OTFS differ fundamentally in their physical-layer signal structures. Unlike OTFS, SC-OFDM(2D) decides a time-frequency grid size according to coherence time and coherence frequency. This approach retains the OFDM standard for pilot design, channel estimation, and equalization, offering a more straightforward and compatible solution. By setting up a 2D grid, SC-OFDM(2D) signals efficiently address delay and Doppler impacts through enhanced time-frequency diversity, a capability that traditional OFDM struggles to achieve. SC-OFDM(1D), by contrast, only provides one-dimensional diversity in the frequency domain without incorporating time diversity. The 2D signal generation process consists of three stages. In the first stage, an $M$-point \ac{DFT} is applied independently to each of the $K$ columns. In the second stage, a $K$-point \ac{IDFT} is performed on each row. Finally, in the third stage, an $N$-point \ac{IFFT} is applied to each column of the newly obtained $M \times K$ matrix, where $M < N$.

\subsection{SC-NOFS(2D)}

As discussed, SC-NOFS(1D) offers many advantages, including improved spectral efficiency, compatibility with OFDM standards, and enhanced frequency diversity. This allows communication systems to achieve higher data rates using OFDM in multipath frequency-selective channels. However, SC-NOFS(1D) is not specifically designed to handle delay-Doppler channels, which are common in applications such as high-speed trains and satellite communications, where these challenging effects are prevalent. Inspired by the SC-OFDM(2D) framework, we propose to introduce both frequency and time diversity into SC-NOFS(2D) by applying 2D precoding in both time and frequency dimensions as illustrated in Fig. \ref{Fig:SC_NOFS_comparison_other_signals}. Enhancing frequency diversity helps address the delay-domain effects, while enhancing time diversity mitigates Doppler-domain effects. This 2D precoding structure enables SC-NOFS(2D) to perform effectively in delay-Doppler channels, expanding its application to environments with complex channel conditions. Similar to SC-OFDM(2D), the 2D signal generation process for SC-NOFS(2D) consists of three stages. In the first stage, an $Q$-point \ac{NOFST} is applied independently to each of the $K$ columns. In the second stage, a $K$-point \ac{IDFT} is performed on each row. Finally, in the third stage, an $N$-point \ac{IFFT} is applied to each column of the newly obtained $Q \times K$ matrix, where $Q < M < N$.

\section{Performance Comparisons}

\begin{figure}[t]
\begin{center}
\includegraphics[scale=0.68]{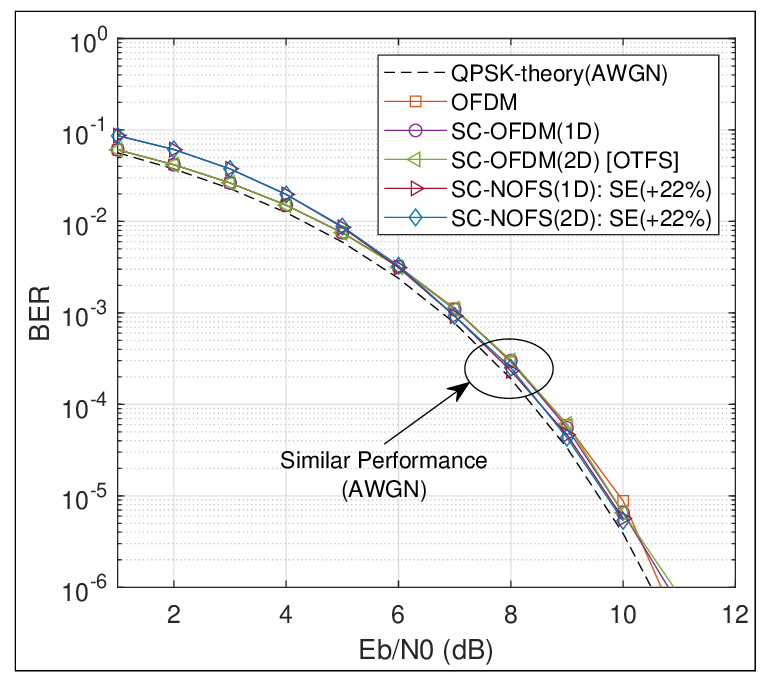}
\end{center}
\caption{{Performance comparison of orthogonal waveforms (OFDM, SC-OFDM(1D), SC-OFDM(2D) [OTFS]) and non-orthogonal waveforms (SC-NOFS(1D), SC-NOFS(2D)) in AWGN channel. Q=492, M=600, N=1024, CP=72. The compression ratio for SC-NOFS(1D,2D) is 0.82 (492/600), the spectral efficiency is increased by approximately 22\%, calculated as (1-0.82)/0.82. SC-OFDM(2D) and OTFS share the same signal structure based on 2D precoding. Their performance is expected to be equivalent in AWGN channel.}}
\label{Fig:BER_NOFS_OTFS_1D_2D_AWGN}
\end{figure}

\begin{figure}[t]
\begin{center}
\includegraphics[scale=0.68]{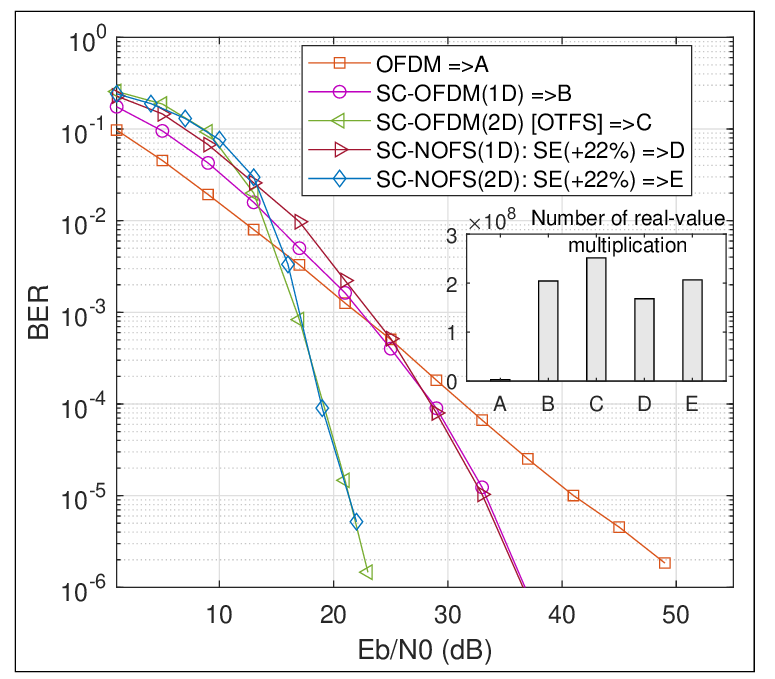}
\end{center}
\caption{{Performance comparison of orthogonal waveforms (OFDM, SC-OFDM(1D), SC-OFDM(2D), {OTFS}) and non-orthogonal waveforms (SC-NOFS(1D), SC-NOFS(2D)) in time-variant multipath frequency selective channel. Q=492, M=600, N=1024, CP=72. The compression ratio for SC-NOFS(1D,2D) is 0.82 (492/600), the spectral efficiency is increased by approximately 22\%, calculated as (1-0.82)/0.82. SC-OFDM(2D) and OTFS share the same signal structure based on 2D precoding. Their performance is expected to be equivalent in scenarios where both waveforms can achieve accurate channel estimation.}}
\label{Fig:BER_NOFS_OTFS_1D_2D_channel_random}
\end{figure}

\begin{figure*}[t!]
\begin{center}
\includegraphics[scale=0.49]{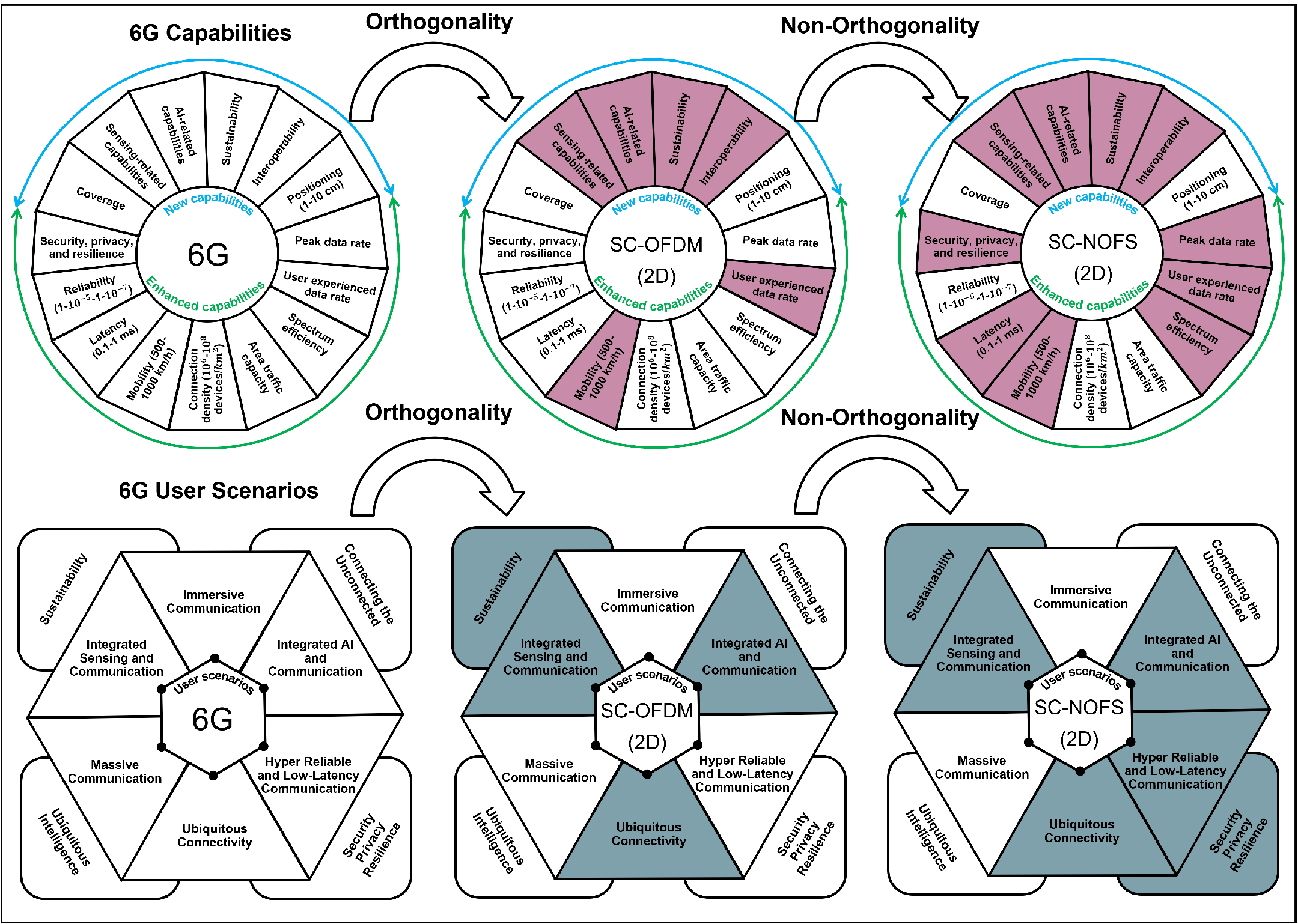}
\end{center}
\caption{{Side-by-side comparison of SC-OFDM (orthogonal) and SC-NOFS (non-orthogonal) waveforms for 6G, evaluated across new capabilities, enhanced capabilities, and user scenarios. Colored blocks indicate the functions supported by SC-OFDM (2D) or SC-NOFS (2D), while uncolored blocks denote unsupported features.}}
\label{Fig:SC_NOFS_comparison_standard_diagram}
\end{figure*}

Before analyzing the impact of multipath channel effects on various signal waveforms, it is crucial to first evaluate their performance in a pure \ac{AWGN} channel. In Fig. \ref{Fig:BER_NOFS_OTFS_1D_2D_AWGN}, we test all waveform candidates under AWGN distortion alone, where each signal achieves performance similar to OFDM. However, when time-variant multipath channel effects are introduced, as shown in Fig. \ref{Fig:BER_NOFS_OTFS_1D_2D_channel_random}, the results vary significantly. This demonstrates that achieving OFDM-like performance in AWGN does not guarantee comparable performance in time-variant multipath channels.

It is worth noting that OTFS employs a non-standard pilot allocation scheme, whereas SC-OFDM(2D) follows the 3GPP-standard pilot structure. Despite these differences in pilot design, both OTFS and SC-OFDM(2D) share the same underlying signal generation structure based on 2D precoding. Therefore, in scenarios where both waveforms can achieve accurate channel estimation, their performance is expected to be equivalent.

Fig. \ref{Fig:BER_NOFS_OTFS_1D_2D_channel_random} presents a comparison of different signal waveforms under a highly frequency-selective and time-variant channel, one of the most challenging wireless environments. The chosen channel model provides a robust test of each waveform's resilience to delay-Doppler impacts. It is noted that the benefits of 2D precoding in SC-NOFS for delay and Doppler resilience, spectral efficiency improvement, and its backward compatibility are not dependent on this specific model alone. It is also possible to use any other channel models for this testing, provided its delay-Doppler effects fall within the coherence time and bandwidth of the waveform. The channel's \ac{PDP} is initialized as ${h(t)}=0.8765\delta(t)-0.2279\delta(t-T_s)+0.1315\delta(t-4T_s)-0.4032e^{\frac{j\pi}{2}}\delta(t-7T_s)$, where $T_s$ indicates the duration of a time sample. To model a time-variant channel, the amplitude for each $\delta(\cdot)$ term varies and follows a Rayleigh distribution. 

In current 4G/5G standards, pilot symbols are typically inserted and distributed every 0.5 ms, which corresponds to the duration of half a sub-frame. During this period, the channel characteristics are assumed to remain static. Based on this, we design a time-variant channel model for the simulation, where the \ac{PDP} changes every 0.5 ms. Specifically, the amplitude for each path in $h(t)$ varies every 0.5 ms. This allows the distributed pilot symbols within a half sub-frame to accurately estimate the rapidly varying channel conditions using the conventional one-tap channel estimation method.

The Doppler frequency can be calculated using $f_m = 0.423/{T_c}$ \cite{book_channel_coherence_time}. With ${T_c} = 0.5$ ms, this yields a Doppler frequency of 846 Hz, computed as $0.423 / (0.5 \times 10^{-3})$. The relationship between Doppler frequency and the velocity of a communication target is given by $f_m = v / \lambda = v \cdotp f_{RF} / c$, where $v$ represents the target's velocity, $\lambda$ is the signal wavelength, $f_{RF}$ is the carrier frequency, and $c$ is the speed of light. For a carrier frequency of 2.4 GHz, the target's velocity corresponds to 380 km/s. Accordingly, simulations are conducted using this channel model, which assumes a Doppler frequency of 846 Hz and a target moving at 380 km/h.

Figure \ref{Fig:BER_NOFS_OTFS_1D_2D_channel_random} illustrates the effectiveness of the frequency-domain one-tap channel estimation and equalization method through BER analysis in the time-varying multipath channel. It is important to highlight that the channel estimation approach for the proposed SC-NOFS and SC-OFDM align with the conventional OFDM principle, ensuring seamless integration into existing standards. It is apparent from the results that OFDM achieves the worst performance due to its vulnerability in time-variant frequency selective channel effects. In contrast, SC-OFDM(1D), while achieving similar performance to OFDM in AWGN channels, outperforms OFDM under frequency-selective conditions because of its frequency diversity capabilities. SC-OFDM(2D) further improves performance by leveraging two-dimensional time-frequency diversity, effectively addressing delay-Doppler effects. It is noted that OTFS can achieve the same performance with SC-OFDM(2D) in scenarios where both waveforms are able to perform accurate channel estimation. The proposed SC-NOFS(1D) exhibits performance comparable to SC-OFDM(1D) due to the enhanced frequency diversity achieved through first-stage signal precoding. Similarly, SC-NOFS(2D) matches the performance of SC-OFDM(2D), benefiting from two-dimensional time-frequency precoding. Importantly, both SC-NOFS(1D) and SC-NOFS(2D) incorporate non-orthogonality features, which provide improved spectral efficiency compared to SC-OFDM(1D) and SC-OFDM(2D). Moreover, as an OFDM-based waveform compatible with 5G standards, SC-NOFS enables seamless system integration into existing communication systems.

\section{Computational Complexity}
The computational complexity of each waveform is analyzed based on the signal processing principles illustrated in Fig. \ref{Fig:SC_NOFS_comparison_other_signals}. In the case of OFDM, generating a frame requires performing $K$ instances of $N$-point IFFT operations. For SC-OFDM (1D), frame generation involves $K$ instances of $M$-point DFT followed by $K$ instances of $N$-point IFFT. Similarly, SC-NOFS (1D) frame generation consists of $K$ instances of $Q$-point NOFST followed by $K$ instances of $N$-point IFFT. For SC-OFDM (2D), the process requires $K$ instances of $M$-point DFT, $M$ instances of $K$-point IDFT, and $K$ instances of $N$-point IFFT. This computational structure is equivalent to that used in OTFS signal generation. Meanwhile, SC-NOFS (2D) frame generation involves $K$ instances of $Q$-point NOFST, $Q$ instances of $K$-point IDFT, and $K$ instances of $N$-point IFFT.

Since NOFS-based signals are implemented using neural network-based methods with real-valued weight matrices, the number of real-valued operations is calculated for each transform. Specifically, an $N$-point IFFT requires $(2N)log_2{N}$ real multiplications and $(3N)log_2{N}$ real additions. An $M$-point DFT demands $4M^2$ real multiplications and $4M^2-2M$ real additions, while a $K$-point IDFT requires $4K^2$ real multiplications and $4K^2-2K$ real additions. Additionally, a $Q$-point NOFST involves $4QM$ real multiplications and $4QM-2Q$ real additions.

In this work, the parameters are configured as $Q$=492, $M$=600, $N$=1024, and $K$=140. Given that multiplication operations dominate hardware resource utilization, the  computational complexity across waveforms is summarized in terms of the number of real-valued multiplications, as depicted in Fig. \ref{Fig:BER_NOFS_OTFS_1D_2D_channel_random}. This analysis reveals key performance-complexity trade-offs among the candidate waveforms. OFDM exhibits the lowest computational complexity but suffers from limited spectral efficiency and high susceptibility to delay-Doppler distortions. In contrast, SC-OFDM (2D), equivalent to OTFS, provides delay-Doppler resilience at the cost of the highest computational complexity. However, OTFS further increases receiver-side complexity due to non-standard channel estimation, equalization, and detection processes. SC-NOFS (2D) presents a balance, with complexity higher than OFDM, comparable to SC-OFDM (1D), yet lower than SC-OFDM (2D), while offering additional benefits in spectral efficiency and delay-Doppler resilience.

\section{Potentials for 6G Standard}

Traditional OFDM and SC-OFDM(1D) are likely to remain competitive in 6G due to their simple signal processing and high spectral and power efficiency. However, these waveforms may struggle to meet the demands of 6G, which requires advanced capabilities and support for new application scenarios, as shown in Fig. \ref{Fig:SC_NOFS_1G_6G_development}. While SC-OFDM(1D) offers frequency domain diversity, it lacks mechanisms to effectively handle delay or Doppler effects. OFDM, as a basic signal format, also lacks built-in methods to mitigate these challenges. Therefore, as seen in the comparison in Fig. \ref{Fig:SC_NOFS_comparison_standard_diagram}, both OFDM and SC-OFDM(1D) are excluded. In contrast, SC-OFDM(2D), with its orthogonal 2D signal structure, demonstrates strong resilience against delay-Doppler effects and benefits from enhanced time-frequency diversity.

In this comparison, SC-OFDM(2D) is chosen as the representative orthogonal signal waveform for 6G. Although NOFS demonstrates a substantial 150\% improvement in spectral efficiency, it is incompatible with existing 3GPP standards, particularly in terms of the over-the-air interface, channel estimation, and equalization. Consequently, NOFS is excluded from this analysis. To address these challenges, SC-NOFS has been introduced as an OFDM-compatible signal waveform, available in both 1D and 2D formats. As outlined in Section \ref{sec:potential_6G_waveforms}, SC-NOFS(1D) serves as a subset of SC-NOFS(2D), with the latter incorporating all the benefits of the 1D format while providing enhanced robustness against delay-Doppler effects. Therefore, the comparison in Fig. \ref{Fig:SC_NOFS_comparison_standard_diagram} focuses on two signal formats: the orthogonal SC-OFDM(2D) and the non-orthogonal SC-NOFS(2D).

Both SC-OFDM(2D) and SC-NOFS(2D) signals are backward compatible with OFDM and thus they both have the capability `Interoperability'. Waveforms like SC-NOFS that preserve OFDM-compatible structures minimize the need for hardware upgrade, which reduces electronic waste and capital expenditure, leading to `Sustainability'. Both signals utilize multicarrier structures, where each sub-carrier can function as a neural network cell, enabling `AI-related capabilities', while variations in each sub-carrier can be leveraged for `sensing-related capabilities'. Additionally, their robustness to delay-Doppler effects makes them ideal for scenarios requiring high `Mobility' and `User Experienced Data Rate'. However, SC-NOFS, with its non-orthogonal characteristics, introduces additional advantages. Its data compression capability improves `Spectral Efficiency' and can either enhance the `Peak Data Rate' or reduce `Latency'. {The non-orthogonal multicarrier structure introduces intentional self-interference, which complicates signal interception and demodulation by eavesdroppers, thereby enhancing the system’s `Security, Privacy, and Resilience' capabilities.} In terms of user scenarios, both signals support `Integrated Sensing and Communication', `Integrated AI and Communication', and `Ubiquitous Connectivity', aligning with the `Sustainability' goal. SC-NOFS can support an additional user scenario in `Hyper Reliable and Low-Latency Communication' further addressing the `Security, Privacy, and Resilience' objective. Thus, by integrating non-orthogonality into an orthogonal framework, SC-NOFS provides more advantages compared to SC-OFDM.

\balance
\section{Discussion and Future Work}

This study evaluates and compares the potential of orthogonal SC-OFDM(1D,2D) and non-orthogonal SC-NOFS(1D,2D) signal waveforms for 6G communication systems. Both waveforms provide significant advancements over traditional OFDM, particularly in addressing delay-Doppler effects, and are backward-compatible with current OFDM standards. SC-NOFS(2D) stands out with its non-orthogonal architecture, delivering enhanced spectral efficiency, reduced latency, and strengthened security through self-interference management. These features position SC-NOFS(2D) as a superior option for supporting critical 6G capabilities, including hyper-reliable low-latency communication, AI integration, and ubiquitous connectivity. While SC-OFDM(2D) remains competitive in traditional applications, SC-NOFS(2D) offers a more robust and adaptable solution, making it well-suited for next-generation networks and a wide range of emerging use cases.

Despite the potentials of SC-NOFS(2D) for future standard integration, it faces a challenge in terms of increased precoding complexity compared to conventional OFDM, which does not require a precoding process. The two-dimensional time-frequency precoding employed in SC-NOFS(2D) introduces additional computational overhead due to the additional transforms needed to process both time and frequency domains. This challenge is not unique to SC-NOFS(2D); it also exists in SC-OFDM(1D,2D) and SC-NOFS(1D), as all these waveforms involve additional precoding steps. Addressing these complexity challenges will require advancements in precoding algorithms and further waveform optimization techniques to make SC-NOFS(2D) a viable and efficient choice for future 6G communications.

\bibliographystyle{IEEEtran}
\bibliography{WCM_Ref}

@ARTICLE{Anderson2013, 
	author={Anderson, J.B. and Rusek, F. and \"{O}wall, V.}, 
	journal={Proceedings of the IEEE}, 
	title={{Faster-than-Nyquist signaling}}, 
	year={2013}, 
	month={Aug.},
	volume={101}, 
	number={8}, 
	pages={1817-1830},  
	doi={10.1109/JPROC.2012.2233451}, 
	ISSN={0018-9219},}

@ARTICLE{FBMC2011_magazine,
	author={Farhang-Boroujeny, B.},
	journal={IEEE Signal Processing Magazine},
	title="{OFDM} Versus Filter Bank Multicarrier",
	year={2011},
	volume={28},
	number={3},
	pages={92-112},
	month={May},}

@ARTICLE{GFDM_trans,
	author={Michailow, N. and Matthe, M. and Gaspar, I.S. and Caldevilla, A.N. and Mendes, L.L. and Festag, A. and Fettweis, G.},
	journal={IEEE Transactions on Communications},
	title={Generalized Frequency Division Multiplexing for 5th Generation Cellular Networks},
	year={2014},
	month={Sep.},
	volume={62},
	number={9},
	pages={3045-3061},}

@INPROCEEDINGS{UFMC2013,
		author={Vakilian, V. and Wild, T. and Schaich, F. and Ten Brink, S. and Frigon, J.-F.},
		booktitle={IEEE Globecom Workshops},
		title="Universal-filtered multi-carrier technique for wireless systems beyond {LTE}",
		year={2013},
		month={Dec.},
		pages={223-228},}

@ARTICLE{TongyangTVT2017,
	author={T. Xu and I. Darwazeh},
	journal={IEEE Transactions on Vehicular Technology},
	title={Transmission Experiment of Bandwidth Compressed Carrier Aggregation in a Realistic Fading Channel},
	year={2017},
	volume={66},
	number={5},
	pages={4087-4097},
	month={May},}

@ARTICLE{Tongyang_JIOT_2018_double_device,
	author={{T. Xu and I. Darwazeh}},
	journal={IEEE Internet of Things Journal}, 
	title={Non-Orthogonal Narrowband Internet of Things: A Design for Saving Bandwidth and Doubling the Number of Connected Devices}, 
	year={2018},
	month={Jun.},
	volume={5},
	number={3},
	pages={2120-2129},
}

@book{Erik_book_5G,
	author    = {Dahlman, Erik. and Parkvall, Stefan. and Sk\"{o}ld, Johan.},
	title     = "{5G NR}: The Next Generation Wireless Access Technology",
	publisher = "Academic Press",
	year      = "2018",
	month={Aug.},
}

@ARTICLE{6G_operator_CM_2024,
	author={Na, Minsoo and Lee, Jaehyun and Choi, Giwan and Yu, Takki and Choi, Jeongsik and Lee, Jinyoung and Bahk, Saewoong},
	journal={IEEE Communications Magazine}, 
	title="Operator's Perspective on {6G: 6G} Services, Vision, and Spectrum", 
	year={2024},
	month={Aug.},
	volume={62},
	number={8},
	pages={178-184},}

@misc{6G_IMT2030,
	author = {{International Telecommunication Union}},
	title = {{IMT towards 2030 and beyond}},
	month = {Jun.},
	year = "2023",
	howpublished={{https://www.itu.int/en/ITU-R/study-groups/rsg5/rwp5d/imt-2030/Pages/default.aspx}},
}

@ARTICLE{Tongyang_Nature_2023,
	author={Xu, T. and Darwazeh, I.},
	journal={Nature Communications Engineering },
	title="Identification and practical validation of spectrally efficient non-orthogonal frequency shaping waveform",
	year={2023},
	month={Aug.},
	volume={2},
	number={58},
}

@ARTICLE{Tongyang_JIOT_2024_SC_OFDM,
	author={Xu, Tongyang and Li, Shuangyang and Yuan, Jinhong},
	journal={IEEE Internet of Things Journal}, 
	title="{OFDM}-Standard Compatible {SC-NOFS} Waveforms for Low-Latency and Jitter-Tolerance Industrial IoT Communications", 
	year={2024},
	month={Aug.},
	volume={11},
	number={16},
	pages={26901-26915},}

@ARTICLE{GreenCom_TGCN_2022,  
	author={Thompson, John S. and Fletcher, Simon and Friderikos, Vasilis and Gao, Yue and Hanzo, Lajos and Reza Nakhai, Mohammad and O'Farrell, Timothy and Wells, Patricia D.},  
	journal={IEEE Transactions on Green Communications and Networking},   
	title={Editorial A Decade of Green Radio and the Path to ``Net Zero'': A United Kingdom Perspective},   
	year={2022},  
	month={Jun.},
	volume={6},  
	number={2},  
	pages={657-664}, 
}

@ARTICLE{6G_green_CM_2023,
	author={Hoßfeld, Tobias and Varela, Martín and Skorin-Kapov, Lea and Heegaard, Poul E.},
	journal={IEEE Communications Magazine}, 
	title="A Greener Experience: Trade-Offs between {QoE} and {CO2} Emissions in Today's and {6G} Networks", 
	year={2023},
	month={Sep.},
	volume={61},
	number={9},
	pages={178-184},}

@book{book_channel_coherence_time,
	author = "Sklar, Bernard and Harris, Fred",
	title = "Digital communications: fundamentals and applications",
	publisher ="London: Pearson",
	edition   = "3rd",
	year = "2021",
	month={Jan.},
}

@INPROCEEDINGS{OTFS_WCNC2017,  
	author={Hadani, R. and Rakib, S. and Tsatsanis, M. and Monk, A. and Goldsmith, A. J. and Molisch, A. F. and Calderbank, R.},  
	booktitle={2017 IEEE Wireless Communications and Networking Conference (WCNC)},   
	title={Orthogonal Time Frequency Space Modulation},   
	year={2017},  
	month={Mar.},
	volume={},  number={},  pages={1-6}, 
}

\end{document}